\documentclass[12pt,preprint]{aastex}

\shorttitle{eruptive and confined solar flares} \shortauthors{Li et
al.}

\begin{document}

\title{Magnetic Flux and Magnetic Non-potentiality of Active Regions in Eruptive and Confined Solar Flares}

\author{Ting Li\altaffilmark{1,2}, Anqin Chen\altaffilmark{3}, Yijun Hou\altaffilmark{1,2}, Astrid
M. Veronig\altaffilmark{4}, Shuhong Yang\altaffilmark{1,2} \& Jun
Zhang\altaffilmark{5}}

\altaffiltext{1}{CAS Key Laboratory of Solar Activity, National
Astronomical Observatories, Chinese Academy of Sciences, Beijing
100101, China; liting@nao.cas.cn} \altaffiltext{2}{School of
Astronomy and Space Science, University of Chinese Academy of
Sciences, Beijing 100049, China} \altaffiltext{3}{Key Laboratory of
Space Weather, National Center for Space Weather, China
Meteorological Administration, Beijing 100081, China}
\altaffiltext{4}{Institute of Physics \& Kanzelh\"ohe Observatory
for Solar and Environmental Research, University of Graz, A-8010
Graz, Austria} \altaffiltext{5}{School of Physics and Materials
Science, Anhui University, Hefei 230601, China}

\begin{abstract}

With the aim of understanding how the magnetic properties of active
regions (ARs) control the eruptive character of solar flares, we
analyze 719 flares of Geostationary Operational Environmental
Satellite (GOES) class $\geq$C5.0 during 2010$-$2019. We carry out
the first statistical study that investigates the flare-coronal mass
ejections (CMEs) association rate as function of the flare intensity
and the AR characteristics that produces the flare, in terms of its
total unsigned magnetic flux ($\Phi$$_{AR}$). Our results show that
the slope of the flare-CME association rate with flare intensity
reveals a steep monotonic decrease with $\Phi$$_{AR}$. This means
that flares of the same GOES class but originating from an AR of
larger $\Phi$$_{AR}$, are much more likely confined. Based on an AR
flux as high as 1.0$\times$$10^{24}$ Mx for solar-type stars, we
estimate that the CME association rate in X100-class ``superflares"
is no more than 50\%. For a sample of 132 flares $\geq$M2.0 class,
we measure three non-potential parameters including the length of
steep gradient polarity inversion line (L$_{SGPIL}$), the total
photospheric free magnetic energy (E$_{free}$) and the area with
large shear angle (A$_{\Psi}$). We find that confined flares tend to
have larger values of L$_{SGPIL}$, E$_{free}$ and A$_{\Psi}$
compared to eruptive flares. Each non-potential parameter shows a
moderate positive correlation with $\Phi$$_{AR}$. Our results imply
that $\Phi$$_{AR}$ is a decisive quantity describing the eruptive
character of a flare, as it provides a global parameter relating to
the strength of the background field confinement.

\end{abstract}



\section{Introduction}

Solar flares and coronal mass ejections (CMEs) are the most
catastrophic phenomena in the present solar system, driven by a
sudden release of magnetic energy stored in the solar corona. Large
solar flares are often, but not always, associated with CMEs. We dub
flares with a CME ``eruptive" and flares without a CME ``confined"
(Moore et al. 2001). There are two main factors that are considered
to determine whether a flare event is CME-eruptive or not. One
factor is the constraining effect of the background magnetic field
overlying the flaring region, i.e., the strength of magnetic field
or its decay with height (Wang \& Zhang 2007; Yang et al. 2014;
Thalmann et al. 2015; Baumgartner et al. 2018). Li et al. (2020)
analyzed 322 large flares and found that the flare-CME association
rate decreases with the increasing magnetic flux of the active
region (AR) that produces the flare, implying that large magnetic
flux generally tends to confine eruptions.

Another factor determining the eruptive character of solar flares is
thought to be related to magnetic complexity and non-potentiality of
ARs (Sun et al. 2015; Liu et al. 2016; Jing et al. 2018; Duan et al.
2019), such as free magnetic energy, relative helicity, magnetic
twists, etc. Numerous statistical studies have revealed that strong
flares mostly occur in ARs with a complex configuration and high
non-potentiality of magnetic fields (Mayfield \& Lawrence 1985;
Falconer et al. 2002; Chen \& Wang 2012, 2020; Su et al. 2014).
However, there are only a few statistical studies focusing on
magnetic non-potential measures in confined versus eruptive flares
(Nindos \& Andrews 2004; Cui et al. 2018; Vasantharaju et al. 2018).
Nindos \& Andrews (2004) and Gupta et al. (2021) found that in a
statistical sense the pre-flare coronal magnetic helicity of ARs
producing confined large flares is smaller than that of ARs
producing eruptive large flares. Bobra \& Ilonidis (2016) used
machine-learning algorithms to predict CME productivity and found
that the ``intensive" parameters (those not scaling with the AR
size) distinguish between eruptive and confined flares. Until now,
the key non-potential parameters of ARs governing the eruptive
character of solar flares are still unknown based on statistical
results. Moreover, the access to open flux (open to the
interplanetary space) is also thought to influence whether an
X-class flare is likely to be eruptive (DeRosa \& Barnes 2018).

In this Letter, we carry out the first statistical study that
investigates the flare-CME association rate R as a function of the
AR characteristics that produces the flare, in terms of its total
magnetic flux ($\Phi$$_{AR}$). Our findings reveal clear differences
of R, with the slope of R as function of flare intensity being a
monotonically decreasing function of $\Phi$$_{AR}$.  This result has
important implications for the prediction of CMEs occurring in
association with large flares as well as for the solar-stellar
connection, where the solar flare-CME association rates are used to
estimate stellar CME occurrence frequencies. Moreover, we also find
the distinct differences of non-potential parameters characterizing
ARs in eruptive and confined large flares.

\section{Observational Data and Analysis}
\subsection{Data and Event Selection}

We check for the soft X-ray (SXR) flare catalog recorded by the
Geostationary Operational Environmental Satellite (GOES) system and
select flare events $\geq$C5.0 occurring within $45^{\circ}$ from
the central meridian, from June 2010 to June 2019. A total of 719
events are selected, including 322 M-class (Li et al. 2020) and 397
C-class flares. To determine whether a flare is associated with a
CME, we use the CME
catalog\footnote{\url{https://cdaw.gsfc.nasa.gov/CME\_list/}} of the
Solar and Heliospheric Observatory (SOHO)/Large Angle and
Spectrometric Coronagraph (LASCO; Brueckner et al. 1995). The
observations from the Atmospheric Imaging Assembly (AIA; Lemen et
al. 2012) on board the \emph{Solar Dynamics Observatory}
(\emph{SDO}; Pesnell et al. 2012) and the twin \emph{Solar
Terrestrial Relations Observatory} (\emph{STEREO}; Kaiser et al.
2008; Howard et al. 2008) are also used to help determining the CME
association (see detailed description in Li et al. 2020). Out of
these 719 flares, 251 events are eruptive and 468 are confined (see
Table 1 and the database
FlareC5.0\footnote{\url{http://dx.doi.org/doi:10.12149/101067}}).
For each event, we calculated $\Phi$$_{AR}$ before the flare onset
(within 30 min) by using the available vector magnetograms from
Space-Weather Helioseismic and Magnetic Imager (HMI; Scherrer et al.
2012) AR Patches (SHARP; Bobra et al. 2014). Only pixels that host a
radial component of the magnetic field $|$$B_{r}$$|$$>$100 G are
considered (Kazachenko et al. 2017).

\subsection{Calculation of Magnetic Non-potential Parameters of ARs}

For a subset of 132 flare events $\geq$M2.0 (86 eruptive and 46
confined), we calculated three non-potential parameters before the
flare onset (within 30 min) including the length of
polarity-inversion lines (PILs) with steep horizontal magnetic
gradient (L$_{SGPIL}$), the total photospheric free magnetic energy
(E$_{free}$) and the area with strong magnetic shear (A$_{\Psi}$).
Magnetic PILs mark the separation between positive and negative
magnetic flux in the photosphere of ARs. Properties of PILs in ARs
have been found to be strongly correlated to solar flare and CME
occurrences (Falconer et al. 2002; Vasantharaju et al. 2018;
Kontogiannis et al. 2019; Wang et al. 2020). High-gradient,
strong-field PILs are proxies of (near-)photospheric compact
electrical currents and the occurrence of major flares was often
associated with the emergence of flux with high-gradient,
strong-field PILs (Schrijver 2007; Toriumi \& Wang 2019). According
to the method of Chen \& Wang (2012), we measured the length of the
PILs with a steep horizontal magnetic gradient ($\geq$ 300 G
Mm$^{-1}$) for each flare event based on the SHARP vector
magnetograms.

The energy that is released during a flare is generally believed to
originate from the free magnetic energy stored primarily in ARs,
which is the amount of magnetic energy in excess of the minimum
energy attributed to the potential field (Molodensky 1974). It was
found that the higher the free magnetic energy stored in an AR, the
larger the size (magnitude) of upcoming flares (Jing et al. 2010; Su
et al. 2014). We use a proxy for the total photospheric free
magnetic energy (Wang et al. 1996; Chen \& Wang 2012), which can be
calculated as
\begin{equation}
E_{free}=\Sigma\rho_{free}dA, \label{eq1}
\end{equation}
where $\rho_{free}$ is a proxy for the density of the free magnetic
energy in the photosphere, defined as
\begin{equation}
\rho_{free}=\frac{|\textbf{B}_{o}-\textbf{B}_{p}|^{2}}{8\pi},
\label{eq1}
\end{equation}
where $\textbf{B}$$_{o}$ and $\textbf{B}$$_{p}$ are the observed and
the potential magnetic fields, respectively. $\textbf{B}$$_{p}$ was
derived from the observed B$_{r}$ component using the Fourier
transform method. $\rho_{free}$ and E$_{free}$ are in units of [erg
cm$^{-3}$] and [erg cm$^{-1}$], respectively. We measured E$_{free}$
by only considering the pixels with $\rho_{free}$
$\geq$4.0$\times$10$^{4}$ erg cm$^{-3}$.

Magnetic shear, defined as the angular difference between the
measured field and the calculated potential field, is another
commonly used parameter in describing the magnetic complexity and
non-potentiality (Wang et al. 1994; Zhang et al. 2007). We measured
the area with shear angle $\geq$80$^{\circ}$, A$_{\Psi}$ (Chen \&
Wang 2012). The shear angle is given by
\begin{equation}
\Psi=\arccos\frac{\textbf{B}_{o}\cdot\textbf{B}_{p}}{|
 B_{o}B_{p}|}.
\label{eq1}
\end{equation}

\section{Statistical Results}

\subsection{Relations of Flare-CME Association Rate with Flare Intensity and Magnetic Flux of ARs}

We investigate the flare-CME association rate R as function of both
the flare class and the total flux of the source AR for 719 flares
($\geq$C5.0 class). Figure 1(a) shows the scatter plot of
$\Phi$$_{AR}$ versus flare peak SXR flux (F$_{SXR}$). Blue (red)
circles are the eruptive (confined) flares. Obviously, when
$\Phi$$_{AR}$ is large enough ($>$1.0$\times$$10^{23}$ Mx; black
dashed line), an overwhelming majority (about 97\%) of flares do not
generate CMEs (57 of 59 flares are confined). Out of the flare
events of $\Phi$$_{AR}$$>$1.0$\times$$10^{23}$ Mx (a total of 59
events), 32 flares occurred in AR 12192, the huge AR known as
flare-rich but CME-poor (Sun et al. 2015), and the fraction is about
54\%. If we remove the events in AR 12192, almost all the events are
confined (26 of 27 flares are confined).

The value of $\Phi$$_{AR}$ for the 719 flares ranges from
8.5$\times$$10^{21}$ Mx to 2.3$\times$$10^{23}$ Mx, and we divide
$\Phi$$_{AR}$ into five subintervals. Figure 1(b) shows the
relations of the association rate R with F$_{SXR}$ within the five
$\Phi$$_{AR}$ subintervals. For each subinterval, R clearly
increases with F$_{SXR}$. Each straight line in Figure 1(b) shows
the linear fit
\begin{equation}
R=\alpha\log{F_{SXR}}+\beta, \label{eq1}
\end{equation}
where R is in percentage and F$_{SXR}$ is in units of W m$^{-2}$.
For the smallest $\Phi$$_{AR}$ subinterval ($\leq$
2.0$\times$$10^{22}$ Mx), the slope $\alpha$ is 113.8$\pm$13.1 and R
reaches 100\% when the flare is $>$M1.3 class (red straight line).
For the subinterval of 2.0$<$$\Phi$$_{AR}$$\leq$3.5$\times$$10^{22}$
Mx, the slope $\alpha$ decreases to 82.0$\pm$10.6 (green straight
line). It can be seen that in ARs with a small $\Phi$$_{AR}$, about
50\% C5.0-class flares have associated CMEs. For the moderate
$\Phi$$_{AR}$ subintervals (blue and orange lines), the slopes
$\alpha$ are 48.2$\pm$5.4 and 38.4$\pm$6.3, respectively. The
Spearman rank order correlation coefficients r$_{s}$ are 0.96 and
0.94, respectively. In ARs with the largest $\Phi$$_{AR}$ ($>$
9.0$\times$$10^{22}$ Mx), R decreases significantly when compared to
subintervals characterized by smaller $\Phi$$_{AR}$ (black straight
line; also see Table 1). The relation between R and F$_{SXR}$ in the
largest $\Phi$$_{AR}$ subinterval is
\begin{equation}
R=(22.9\pm3.8)\log{F_{SXR}}+(125.7\pm17.9). \label{eq1}
\end{equation}
Based on Eq. 5, only 20\% of all M-class flares originating from the
largest ARs have associated CMEs and the rate R reaches about 40\%
for flares $>$X2. Almost all of C5.0-class flares are confined due
to the strong constraining fields in the largest ARs.

Figure 1(c) shows the relation of the slope $\alpha$ with
$\Phi$$_{AR}$. $\Phi$$_{AR}$ is here defined as the mean of the
individual log values in each $\Phi$$_{AR}$ subinterval. The plot
shows that the slope $\alpha$ decreases monotonically with
increasing $\Phi$$_{AR}$. We assume ARs in solar-type stars of
$\Phi$$_{AR}$ $\sim$ 1.0$\times$$10^{24}$ Mx (Maehara et al. 2012;
Shibata et al. 2013). For there are only five known slope values, it
is difficult to fit the plot and make an extrapolation. We use the
slope in the largest solar ARs, i.e., 22.9, minus the average error
estimate of five known slope values ($\sim$7.8, corresponding to the
average error of five diamonds), that is 22.9 minus 7.8 equals 15.1
as the slope $\alpha$ for stellar ARs of $\Phi$$_{AR}$ $\sim$
1.0$\times$$10^{24}$ Mx. We estimate that the slope $\alpha$ might
be no more than 15.1$\pm$7.8 (red circle). If C5.0-class flares are
all confined on solar-type stars (similar to the subinterval of
$\Phi$$_{AR}$ $>$ 9.0$\times$$10^{22}$ Mx), we can extrapolate the
flare-CME association rate for solar-type stars is given as
\begin{equation}
R=(15.1\pm7.8)\log{F_{SXR}}+80.0. \label{eq1}
\end{equation}
Thus, for X100-class superflares in solar-type stars, the estimated
association rate R is no more than 50\%.

\subsection{Magnetic Non-potentiality of ARs in Eruptive and Confined Flares}

We calculate three non-potential parameters for 132 flares
$\geq$M2.0. Figure 2 shows an example of an eruptive X2.2-class
flare occurring on 2011 February 15 in AR NOAA 11158. The SGPIL is
located between two flare ribbons, and the distributions of the
photospheric free energy density $\rho_{free}$ and shear angle
$\Psi$ show similar patterns. In Figure 3, we display the scatter
plots and histograms of the three non-potential measures for the
whole set of 132 flare events. It can be seen that the distributions
of L$_{SGPIL}$ show evident differences between confined and
eruptive cases (Figures 3(a)-(b)). For L$_{SGPIL}$$<$22 Mm (black
dash-dotted line in Figure 3(a)), an overwhelming majority (about
90\%) of flares are eruptive. The log-mean value of L$_{SGPIL}$ for
confined flares is 40.2 Mm (red dotted line in panel (b)), much
larger than that for eruptive events (20.8 Mm, blue dotted line in
panel (b)). The distributions of E$_{free}$ are similar to those of
L$_{SGPIL}$. For E$_{free}$ $<$1.5$\times$$10^{23}$ erg cm$^{-1}$
(black dash-dotted line in Figure 3(c)), 48 in 58 flares are
eruptive. For A$_{\Psi}$$<$60 Mm$^{2}$, about 79\% (30 in 38) flares
are eruptive (Figure 3(e)). The log-mean values of E$_{free}$ and
A$_{\Psi}$ for confined flares are much larger than those for
eruptive events (Figures 3(d) and (f)).

In Figure 4, we investigate the relationship between the three
non-potential parameters and $\Phi$$_{AR}$ in Figure 4. It can be
seen that each two non-potential parameters have strong correlations
at r$_{s}$ $\sim$ 0.69$-$0.85 (Figures 4(a)-(c)). The high
correlation coefficients between L$_{SGPIL}$, E$_{free}$ and
A$_{\Psi}$ imply that ARs with long SGPIL tend to store more free
magnetic energy and have strong shearing. The scatter plot of
L$_{SGPIL}$ with $\Phi$$_{AR}$ illustrates that they have a moderate
correlation at r$_{s}$ $\sim$ 0.4 (Figure 4(d)). About 89\% of the
confined flares occur in ARs with $\Phi$$_{AR}$ $>$
3.5$\times$$10^{22}$ Mx and SGPIL longer than 22 Mm. Moderate
correlations were also obtained for E$_{free}$ versus $\Phi$$_{AR}$
and A$_{\Psi}$ versus $\Phi$$_{AR}$ (Figures 4(e)-4(f)). Their
moderate correlations with $\Phi$$_{AR}$ indicate that there is some
trend that the more magnetic flux of an AR, the higher
non-potentiality of the AR.

\section{Summary and Discussion}

In this study, we have examined 719 flares $\geq$C5.0-class that
were observed on-disk from June 2010 to June 2019. We investigate
for the first time the flare-CME association rate R as function of
both the flare class F$_{SXR}$ and the total flux $\Phi$$_{AR}$ of
the AR that produces the flare. We find that, for each $\Phi$$_{AR}$
subinterval, R clearly increases with F$_{SXR}$, i.e., larger flares
are more likely associated with a CME. This result is in agreement
with previous findings studying CME-flare association rates (Andrews
2003; Yashiro et al. 2006), who reported overall CME associations of
about 60\% for M-class flares and 90\% for X-class flares. However,
what is new and particularly important in our study is that we
considered not only the relation to the intensity of the flare but
also to the characteristics of the AR in terms of its total magnetic
flux. Our results show that the slope of the flare-CME association
rate depends on the total flux of the AR that produces the flare,
and reveals a steep monotonic decrease with $\Phi$$_{AR}$ (Figure
1(c)). This means that flares of the same GOES class but originating
from an AR of larger $\Phi$$_{AR}$, are much more likely confined.
Within the smallest $\Phi$$_{AR}$ subinterval ($\leq$
2.0$\times$$10^{22}$ Mx), all flares $>$M1.3 are all eruptive. On
the other end of the distribution for the largest $\Phi$$_{AR}$
subinterval ($>$ 9.0$\times$$10^{22}$ Mx), only about 20\% M-class
flares have associated CMEs and the association rate R reaches about
40\% for those flares $>$X2.

Our results imply that $\Phi$$_{AR}$ is a key factor determining the
eruptive character of solar flares, consistent with our previous
studies (Li et al. 2020). $\Phi$$_{AR}$ can be considered to be both
a measure of the total flux that is in principle available for
flaring as well as being a measure of the background field
confinement overlying the flaring region. Our findings imply that
the latter is the more important factor here. Large $\Phi$$_{AR}$
means a strong confinement and thus the flare-CME association rate
is relatively low compared to small $\Phi$$_{AR}$. Moreover, based
on solar observations, we can speculate the associate rate R on
solar-type stars by assuming $\Phi$$_{AR}$ of 1.0$\times$$10^{24}$
Mx (Maehara et al. 2012; Shibata et al. 2013). For X100-class
``superflares" on solar-type stars, no more than 50\% flares can
generate stellar CMEs. This may provide an explanation why the
detection of stellar CMEs is rare (e.g. Leitzinger et al. 2014, Vida
et al. 2019, Argiroffi et al. 2019, Moschou et al. 2019, Veronig et
al. 2021), while extrapolating current flare-CME relations to
solar-type stars leads to unphysically high CME rates (Drake et al.
2013; Odert et al. 2017). Our findings provide an important
contribution to revise the flare-CME association rates for
solar-type stars, by including the distinct differences in these
relations in dependence of the AR magnetic flux.

Using HMI vector magnetograms, we also have studied the relation
between the degree of magnetic non-potentiality and the eruptive
character of 132 flares $\geq$M2.0, finding distinct differences
between eruptive and confined flares for all three non-potentiality
parameters derived (Figures 3-4). L$_{SGPIL}$, E$_{free}$ and
A$_{\Psi}$ all give smaller log-mean values for eruptive than
non-eruptive flares. Each non-potential parameter shows a moderate
correlation with $\Phi$$_{AR}$. Our study shows that the three
``extensive" parameters (those scaling with the AR size) can only
discriminate the flares with small non-potential parameters. As seen
in Figure 3, 42 out of 47 events with L$_{SGPIL}$$<$22 Mm are
eruptive. However, for the remanent 85 flares with L$_{SGPIL}$$>$22
Mm, it is difficult to tell whether a flare is accompanied by a CME
or not. The appearance of other two parameters is similar to that of
L$_{SGPIL}$. Only a fraction of flare events can be discriminated
between confined and eruptive events. Thus we speculate that if we
use these ``extensive" parameters to predict the CME productivity,
the True Skill Score (TSS) value is not high. This is consistent
with the study of Bobra \& Ilonidis (2016), who shows that TSS for
predicting the CME productivity is low based on ``extensive"
parameters. They show that ``intensive" parameters can predict the
CME productivity. In our study, we did not calculate ``intensive"
parameters. In our future work, we will consider ``intensive"
parameters in confined and eruptive large flares based on our
database.

In the statistical results of Cui et al. (2018), confined flares
have larger values of $\Phi$$_{AR}$ and the gradient-weighted area
of the polarity inversion region than eruptive flares, which are in
agreement with our results. It was found that large confined flares
tend to occur in large ARs (Toriumi et al. 2017; Li et al. 2020),
and thus L$_{SGPIL}$, E$_{free}$ and A$_{\Psi}$ are larger for
confined than eruptive flares due to their positive correlations
with $\Phi$$_{AR}$. Our results imply that $\Phi$$_{AR}$ is a key
factor in determining whether a flare is eruptive or confined, as it
provides a global parameter relating to the strength of the
background field confinement.

\acknowledgments {This work is supported by the National Key R\&D
Program of China (2019YFA0405000), the B-type Strategic Priority
Program of the Chinese Academy of Sciences (XDB41000000), the
National Natural Science Foundations of China (11773039, 11903050,
12073001, 11790304, 11873059 and 11790300), Key Programs of the
Chinese Academy of Sciences (QYZDJ-SSW-SLH050), the Youth Innovation
Promotion Association of CAS (2014043 and 2017078), Yunnan
Academician Workstation of Wang Jingxiu (No. 202005AF150025) and
NAOC Nebula Talents Program. A. Q. Chen is supported by the
Strategic Priority Program on Space Science, Chinese Academy of
Sciences, Grant No. XDA15350203. Astrid M. Veronig acknowledges the
support by the Austrian Science Fund (FWF): P27292-N20. \emph{SDO}
is a mission of NASA's Living With a Star Program.}

{}
\clearpage

\begin{table*}
\centering \caption{Number of Eruptive and Confined Flares in all
ARs and ARs with largest $\Phi$$_{AR}$ \label{tab1}} \centering
\begin{tabular}{c c c c c c c} 
\hline\hline 
Class & Eruptive\tablenotemark{1} & Confined\tablenotemark{1} &
R\tablenotemark{2} & Eruptive\tablenotemark{3} & Confined\tablenotemark{3} & R\tablenotemark{2} \\ 
\hline 
C & 82 & 315 & 21\% & 1 & 40 & 2\% \\ M & 154 & 147 & 51\% & 6 & 30 & 17\% \\
X & 15 & 6 & 71\% & 1 & 4 & 20\% \\ \hline Total & 251 & 468 & 35\%
& 8 & 74 & 10\%
\\\hline
\end{tabular}
\tablenotetext{1}{For all ARs} \tablenotetext{2}{Flare-CME
association rate} \tablenotetext{3}{For ARs with largest
$\Phi$$_{AR}$ $>$ 9.0$\times$$10^{22}$ Mx}
\end{table*}

\begin{figure}
\centering
\includegraphics
[bb=98 78 455 733,clip,angle=0,scale=0.5]{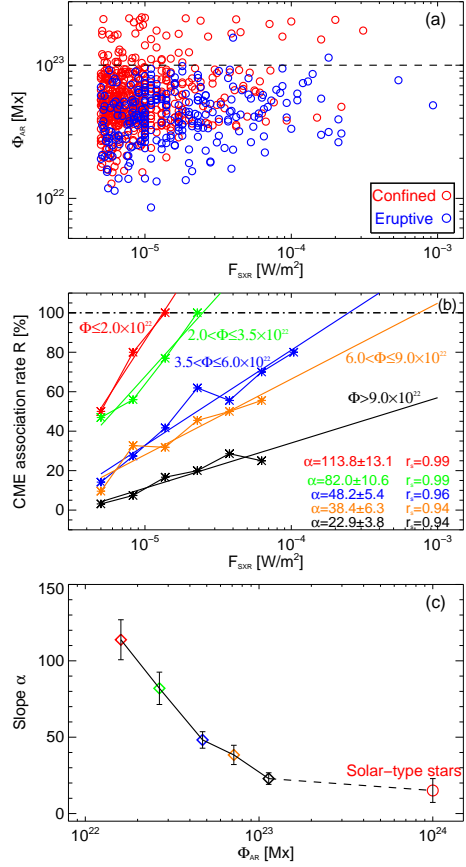}
\caption{Relations of flare-CME association rate (R) with flare peak
soft X-ray flux (F$_{SXR}$) and total unsigned magnetic flux of ARs
($\Phi$$_{AR}$). (a) Scatter plots of $\Phi$$_{AR}$ vs. F$_{SXR}$.
Blue (red) circles are the eruptive (confined) flares ($\geq$
C5.0-class). Black dashed line corresponds to $\Phi$$_{AR}$ of
1.0$\times$$10^{23}$ Mx. (b) Association rate R as a function of
F$_{SXR}$ separately for five different subintervals of
$\Phi$$_{AR}$. The colored straight lines show the results of linear
fitting, and slopes $\alpha$ and Spearman rank order correlation
coefficients r$_{s}$ are shown at the bottom right. (c) Plot of
slope $\alpha$ vs. $\Phi$$_{AR}$ (plotted at the average of the log
values in each subinterval). Colored diamonds denote the slopes
$\alpha$ in five different $\Phi$$_{AR}$ subintervals. The red
circle is the estimated value of slope $\alpha$ (about 15.1) for
solar-type stars by assuming $\Phi$$_{AR}$ of 1.0$\times$$10^{24}$
Mx (Maehara et al. 2012; Shibata et al. 2013). \label{fig1}}
\end{figure}
\clearpage

\begin{figure}
\centering
\includegraphics
[bb=27 253 583 563,clip,angle=0,scale=0.6]{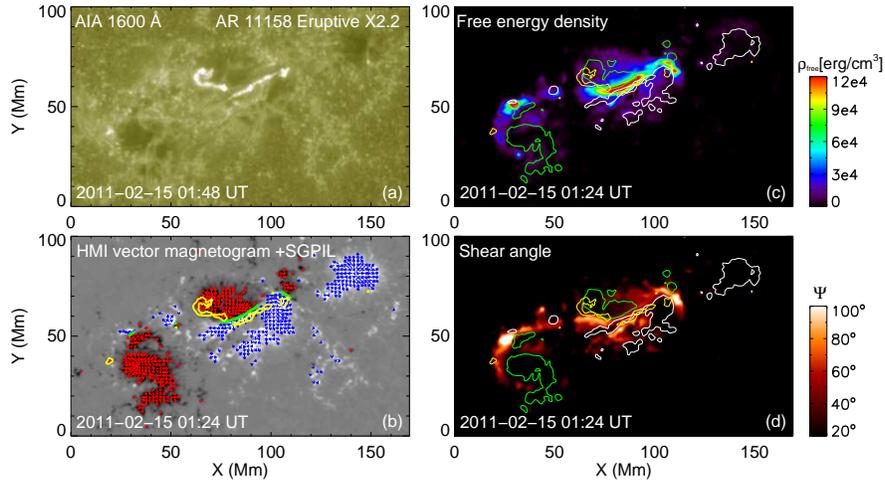}
\caption{Tracing the magnetic polarity inversion line with the steep
horizontal magnetic gradient (SGPIL), photospheric free magnetic
energy density $\rho_{free}$ and the magnetic shear angle $\Psi$ in
AR 11158 on 2011 February 15. (a) SDO/AIA 1600 {\AA} image showing
the flare ribbon brightenings of an eruptive X2.2-class flare. (b)
SDO/HMI vector magnetogram with horizontal magnetic field vectors
(red and blue arrows) overplotted on the B$_{r}$ map. SGPIL ($\geq$
300 G Mm$^{-1}$) is shown as green lines. Yellow contours are the
ribbon brightenings. (c) $\rho_{free}$ distribution with contours of
the B$_{r}$ component and flare ribbon brightenings. The white
(green) contours represent the positive (negative) polarity. (d)
$\Psi$ distribution with contours of the B$_{r}$ component and
ribbon brightenings. \label{fig2}}
\end{figure}
\clearpage

\begin{figure}
\centering
\includegraphics
[bb=26 110 531 703,clip,angle=0,scale=0.6]{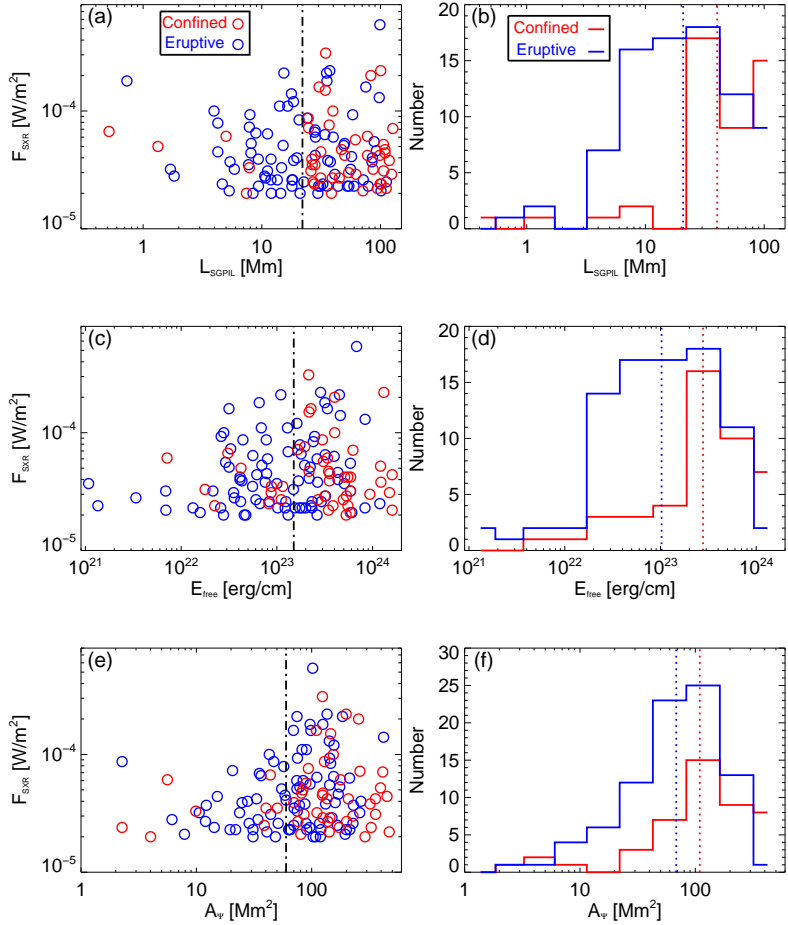}
\caption{Scatter plots and histograms of three different
non-potentiality measures for eruptive (blue) and confined (red)
flares ($\geq$ M2.0-class). Top: scatter plot of F$_{SXR}$ vs. SGPIL
length (L$_{SGPIL}$) and the histogram of L$_{SGPIL}$. Black
dashed-dotted line in panel (a) refers to L$_{SGPIL}$ of 22 Mm.
Dotted vertical lines in panel (b) indicate the means of the log
values. Middle: scatter plot of F$_{SXR}$ vs. the total free
magnetic energy E$_{free}$ and the histogram of E$_{free}$. Black
dashed-dotted line in panel (c) refers to E$_{free}$ of
1.5$\times$$10^{23}$ erg cm$^{-1}$. Bottom: scatter plot of
F$_{SXR}$ vs. the area with strong magnetic shear A$_{\Psi}$ and the
histogram of A$_{\Psi}$. Black dashed-dotted line in panel (e)
refers to A$_{\Psi}$ of 60 Mm$^{2}$. \label{fig3}}
\end{figure}
\clearpage

\begin{figure}
\centering
\includegraphics
[bb=28 139 527 672,clip,angle=0,scale=0.6]{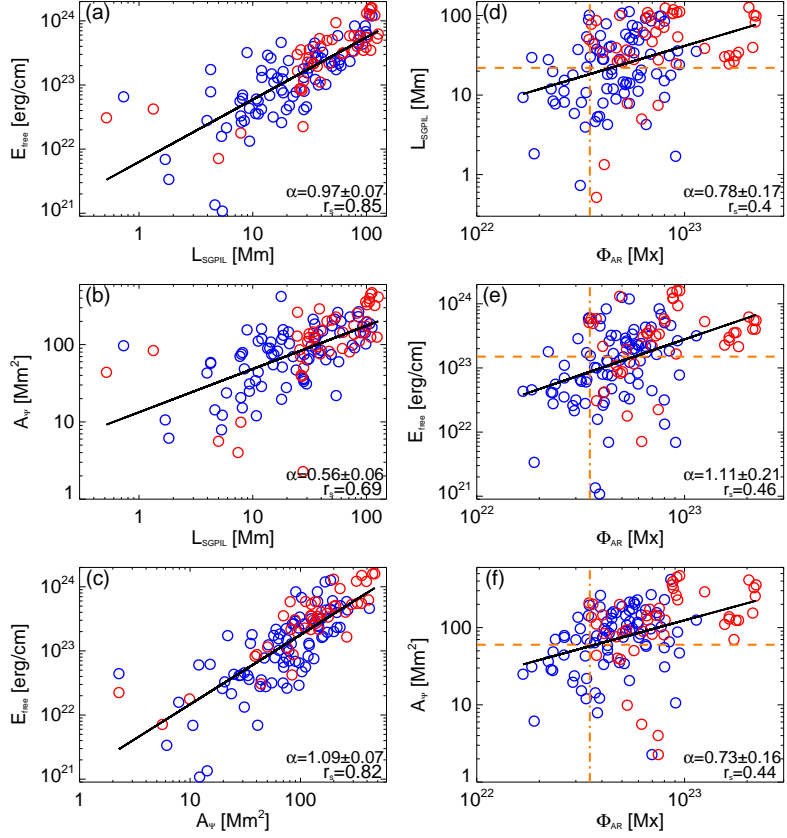} \caption{
Relations between L$_{SGPIL}$, E$_{free}$, A$_{\Psi}$ and
$\Phi$$_{AR}$ for eruptive (blue) and confined (red) flares. The
black solid lines show the results of a linear fitting, and slopes
$\alpha$ and Spearman rank order correlation coefficients r$_{s}$
are shown at the bottom right of each panel. Orange dash-dotted
lines in panels (d)-(f) correspond to $\Phi$$_{AR}$ of
3.5$\times$$10^{22}$ Mx. Orange dashed lines respectively denote
L$_{SGPIL}$ of 22 Mm, E$_{free}$ of 1.5$\times$$10^{23}$ erg
cm$^{-1}$ and A$_{\Psi}$ of 60 Mm$^{2}$. \label{fig4}}
\end{figure}
\clearpage

\end{document}